\documentclass[aps,epsf,preprint]{revtex4}
\usepackage{latexsym}
\usepackage{amsfonts}
\usepackage{graphicx}
\usepackage{setspace}  
\usepackage{epsfig}

\begin{document}

\title{A Physical Mechanism Underlying the Increase of Aqueous
Solubility of Nonpolar Compounds and the Denaturation of Proteins upon
Cooling}

\author{Sergey V. Buldyrev,$^1$ Pradeep Kumar,$^2$ and H. Eugene Stanley$^2$}

\bigskip
\bigskip

\affiliation {$^1$Department of Physics,~Yeshiva
University, 500 West 185th Street,~New York, NY 10033 USA\\
$^2$Center for Polymer Studies and Department of Physics,~Boston
University,~Boston, MA 02215 USA}
\date{19 January 2007, 01:00pm} \pacs{05.40.-a}
\noindent

\begin{abstract}

The increase of aqueous solubility of nonpolar compounds upon cooling
and the cold denaturation of proteins are established experimental
facts. Both phenomena have been hypothesized to be related to
restructuring of the hydrogen bond network of water around small
nonpolar solutes or hydrophobic amino acid side chains. However, an
underlying physical mechanism has yet to be identified.  We assume
the solute particles and the monomers of a polymer interact via a hard
sphere potential. We further assume that the solvent molecules
interact via the two-scale spherically symmetric Jagla potential,
which qualitatively reproduces the anomalies of water, such as
expansion on cooling. We find that this model correctly predicts the
increase in solubility of nonpolar compounds and the swelling of
polymers on cooling.  Our findings are consistent with the possibility
that the presence of two length scales in the Jagla potential---a rigid hard
core and a more flexible soft core---is responsible for both
phenomena.  At low temperatures, the solvent particles prefer to
remain at the soft core distance, leaving enough space for small
nonpolar solutes to enter the solvent thus increasing solubility. We
support this hypothesized mechanism by molecular dynamic simulations.

\end{abstract}

\maketitle
\section{introduction}

The increase of the solubility of nonpolar compounds in water on
lowering the temperature, $T$, is a well known phenomenon
\cite{Zumdahl,Acree84} which recently received attention from the
molecular simulation community \cite{Konrad05,Lenart06,Krouskop06}.  The
existence of closed demixing regions in the temperature-concentration
phase diagrams characterized by the presence of a lower critical solution temperature
(LCST) below which the components mix in all proportions is also well
known for aqueous solutions of polymers and large organic molecules \cite{Acree84} 
and is explained by the difference in the thermal expansion
coefficients of solvent and solute
\cite{Flory64,Flory65,Prigogine57,Van84,Costas84,Patterson70,
Paricaud03}. The same effect, increase of solubility of the hydrophobic
side chains on lowering T, may be relevant for understanding the
remarkable phenomenon of cold denaturation,\cite{pace68,privalov88,privalov13,Jonas97}
a subject of considerable current theoretical \cite{Paschek05,Marques06} and experimental
\cite{Moghaddam05} studies.

The thermodynamic framework of the solubility is well established
\cite{Paricaud03}.  The excess Gibbs free energy $\Delta g$ of
dissolving one mole of solute in $x$ moles of solvent is defined as the
difference between the Gibbs free energies of the solution and the sum
of the Gibbs free energies of the pure solvent and the pure solute at
constant pressure $P$ and temperature $T$,
\begin{equation}
\Delta g=\Delta h-T\Delta s=\Delta u +P\Delta v -T\Delta s,
\label{eq:DeltaG}
\end{equation}
where $\Delta h$, $\Delta u$, $\Delta v$, and $\Delta s$ are the
differences between the molar enthalpy molar potential energy, molar volume
and molar entropy of the solution and pure substances.  The main
contribution to the entropic term $\Delta s\equiv\Delta s_0+ \Delta
s_{\rm mix}$, is the entropy of mixing
\begin{equation}
\Delta s_{\rm mix}(\Phi)=-R\left[\ln\Phi+\left({1\over\Phi}
  -1\right)\ln(1-\Phi)\right], 
\end{equation}
where $\Phi=1/(1+x)$ is the mole fraction of solute, $R \equiv N_A k_B$
is the universal gas constant, $N_A$ is the Avogadro number and $k_B$ is
the Boltzmann constant.  The term T$\Delta s_{\rm mix}$ dominates at
high temperatures and thus favors mixing, but at low temperatures it
becomes insignificant. However, the entropic term is not responsible for
the solubility minimum as function of temperature. According to Le
Chatelier's principle, if solvation is exothermic ($\Delta h <0$) the
solubility increases upon cooling at constant pressure. Conversely, if
$\Delta h>0$, the solubility increases upon heating. Thus the condition
$\Delta h=0$ corresponds to the solubility extreme. Since $\partial
(\Delta g/T) /\partial T\vert_P \equiv -\Delta h/T^2$, where $\Delta
h\equiv \Delta u +P\Delta V$, the maximum in $\Delta g/T$ corresponds to
the point at which $\Delta h=0$, changing its sign from positive to
negative upon cooling.

These thermodynamic arguments were applied by Paschek et al.,
\cite{Paschek04,Paschek05} who showed that for the solution of
Lennard-Jones polymer-chains in TIP5P water the enthalpic term $\Delta
h$ becomes increasingly negative at low temperatures and thus stabilizes
the solution.  The phenomenon of a solubility minimum has been discussed
in terms of hydrogen bond restructuring in water at low temperature,
which leads to the formation of water cages around small nonpolar
molecules, monomers, or protein residues
\cite{frank45,soper06,Errington98}.  Indeed, the $P\Delta v$ term
becomes negative at low temperatures since, because the distances
between water molecules increases upon cooling, there is more space for
small solutes to penetrate the solvent.  The $\Delta u$ term may also
become negative, because introducing small nonpolar solutes inside water
separates water molecules further apart and thus helps water molecules
to form stronger hydrogen bonds.  Without nonpolar particles,
5-coordinated molecules with weaker hydrogen bonds would be observed.

In this paper, we attempt to elucidate the connection of the increase of
solubility upon cooling with the anomalous thermodynamic properties of
pure water.  Recently it was shown
that the spherically-symmetric Jagla ramp potential
(Fig.~\ref{fig:Jagla}) explains anomalous {\it thermodynamic\/} and {\it
dynamic\/} properties of water in terms of the presence of two length
scales: a ``hard core,'' corresponding to the first rigid tetrahedral
shell of the nearest neighbors in water, and a wider ``soft core,''
corresponding to the second (more flexible) shell of neighbors
\cite{Yan05,Xu06b}. We will see that the Jagla potential is capable of
reproducing the {\it solubility\/} properties of water, and based on
this result we will propose a mechanism for the solubility increase of
nonpolar compounds in water on cooling.

Using discrete molecular dynamic (DMD) simulations described in Section
II, we show in Section III that the solubility of hard sphere solutes in
the Jagla solvent has a characteristic minimum as a function of
temperature similar to the minimum of the solubility of small nonpolar
compounds in water. Hard spheres can be used to describe nonpolar
compounds, since the van der Waals interactions among nonpolar compounds
are an order of magnitude weaker than the hydrogen bond interactions
between water molecules and thus can be neglected in assessing the
effect largely caused by the restructuring of hydrogen bonds.  In
Section IV we relate the solubility increase to the behavior of the
excess thermodynamic functions. In Section V, we study the behavior of
the beads-on-a-string homopolymer composed of hard sphere monomers
immersed in the same Jagla solvent.  We show that the radius of gyration
$R_g$ of the polymer has a sharp minimum as a function of temperature at
low constant pressures, indicating that the system may display a LCST at
low pressure.  At higher pressure, the $R_g$ minimum becomes less
pronounced, indicating the increase of solubility at intermediate
temperatures and decrease of solubility at low temperatures.

%\begin{equation}
%U(r) = \left\{
%\begin{array}{ll}
%\infty & r < a\\ U_A+(U_A-U_R)(r-b)/(b-a) & a%b c
%\end{array}\right.
%\label{eq:potential}
%\end{equation}

\section {Methods}

The interaction potential of the Jagla solvent particles $U(r)$ is
characterized by five parameters: the hard core diameter $a$, the soft
core diameter $b$, the range of attractive interactions $c$, the depth
of the attractive ramp $U_A$, and the height of repulsive ramp $U_R$
(Fig.~\ref{fig:Jagla}) \cite{Jagla98}, of which three are independent:
$b/a$, $c/a$, and $U_R/U_A$. The most important of these parameters is
the ratio of the soft core and hard core diameters, $b/a$, which must
be selected close to $r_2/r_1$, where $r_1$ and $r_2$ are the
positions of the first and the second peaks of the oxygen-oxygen
radial distribution function for water.  Following
\cite{Jagla98,Xu06b,Xu05,Xu06a} we select $b/a=1.72$, $c/a=3$,
$U_R/U_A=3.5$. This choice of parameters produces a phase diagram
which qualitatively resembles the water phase diagram with two stable
critical points and a wide region of density anomaly bounded by the
locus of the temperature of maximum density $T_{\rm MD}$
[Fig.~\ref{fig:Jagla}(b)]. The slope of the coexistence line between
the high density liquid (HDL) and the low density liquid (LDL) in this
model is positive---unlike water, for which this slope is negative. In
order to use the DMD algorithm \cite{Alder59,Rapaport78,RAPAPORT}, we
replace the repulsive and attractive ramps with 40 and 8 equal steps,
respectively, as described in \cite{Xu06b} (Fig.~\ref{fig:Jagla}).  We
measure length in units of $a$, time in units of $a\sqrt{m/U_A}$,
where $m$ is the particle mass, density in units of $a^{-3}$, pressure
in units of $U_A/a^3$, and temperature in units of $U_A/k_B$. This
realization of the Jagla model displays a liquid-gas critical point at
$T_{c1}=1.446$, $P_{c1}=0.0417$ $\rho_{c1}=0.102$, and a liquid-liquid
critical point at $T_{c2}=0.375$, $P_{c2}=0.243$
$\rho_{c2}=0.370$~\cite{Xu06b}. We model solute particles by hard
spheres of diameter $d_0$, which we select to be equal to the hard
core diameter of the Jagla solvent $d_0=a$. The dependence of the
solubility on $d_0$ is an important question and will be studied
elsewhere.

\section{Calculation of the Henry Constant}

We first study the effect of pressure and temperature on the solubility
of hard-sphere solutes in the Jagla solvent. In order to do this, we
create a system of N=1400 Jagla particles and 2800 hard spheres in a box
$L_x\times L_y \times L_z$ with periodic boundaries. We fix
$L_x=L_y=15a$, and vary $L_z$ [Fig.~\ref{fig:Jagla}(c)] to maintain
constant pressure and temperature using Berendsen thermostat and
barostat. For $T < T_{c1}$, the mixture of Jagla particles and hard
spheres segregates and the Jagla particles form a slab of liquid
crossing the system perpendicular to the $z$-axis. For each pressure and
temperature, we equilibrate the system for 500 time units and measure
the mole fraction of hard spheres in the narrow slabs perpendicular to
the $z$ axis for another 500 time units.  We find that this
equilibration time is sufficient if the temperature for the two
successive simulations is changed by less then 10\%.  In each slab we
count the numbers $N_J$ of Jagla solvent particles and $N_S$ of hard
sphere solute particles.

We define a slab as a liquid slab if $N_J>N_0$ and as a vapor slab if
$N_J\leq N_0$, where $N_0$ is a temperature and pressure dependent
threshold, such that the liquid and vapor slabs form two continuous
regions covering the entire system with the exception of a few slabs
representing the boundary. To minimize the boundary effects, we exclude
from the liquid phase the slabs whose distance to the closest vapor slab
is less than $6a$. The analogous criterion is applied to determine the
vapor phase.  Finally, we find the mole fraction $\Phi=N_S/(N_S+N_J)$ of
hard spheres in the liquid and vapor phases and compute the Henry
constant $k_H(T,P)=P\Phi_v(T,P)/\Phi_\ell (T,P)$, where $\Phi_v$ and
$\Phi_\ell $ are the average mole fractions of hard spheres in the vapor
and liquid phases respectively.

\begin{figure}[htb]
\centerline{
\includegraphics[angle=270,width=8cm]{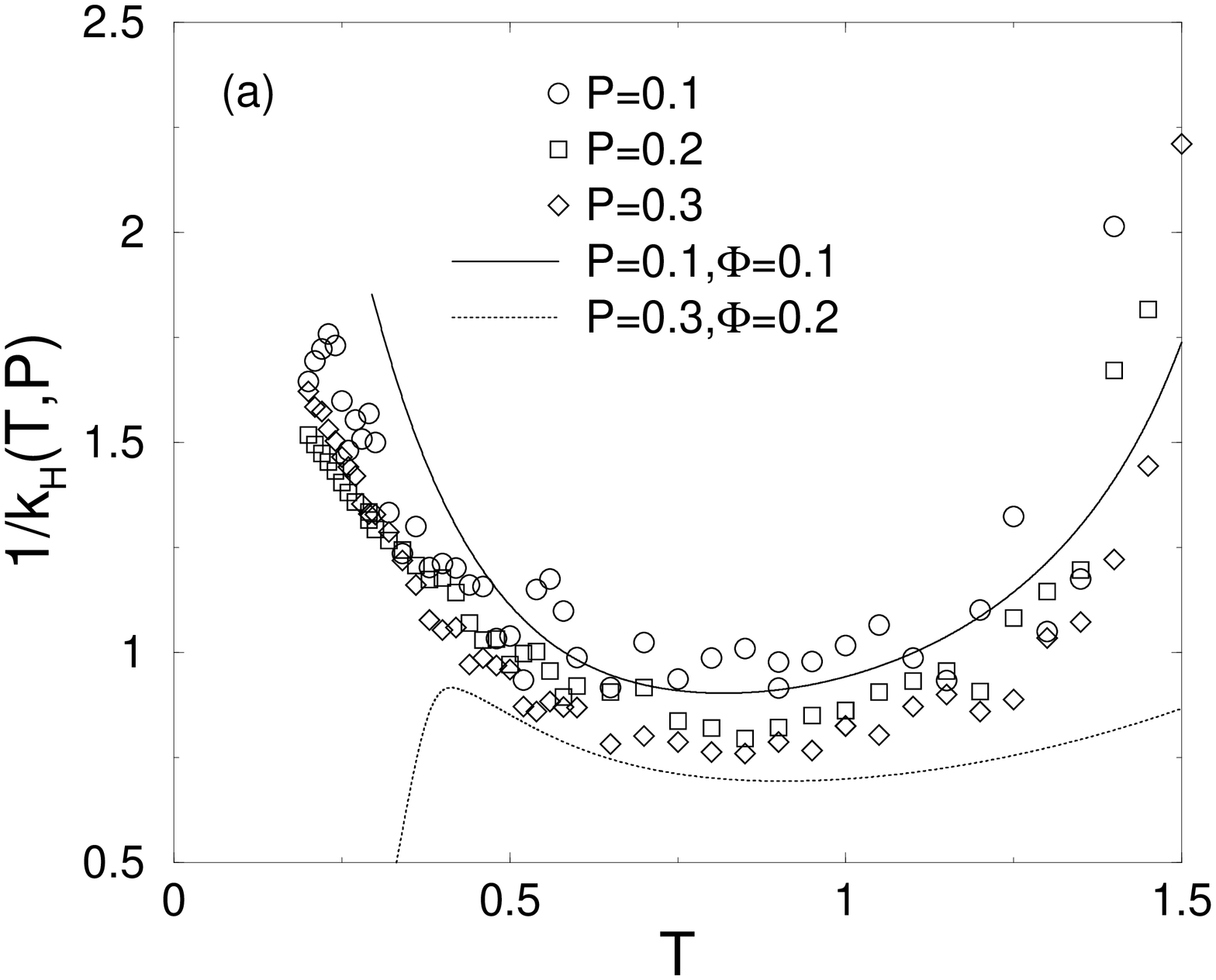}
\includegraphics[angle=270,width=8cm]{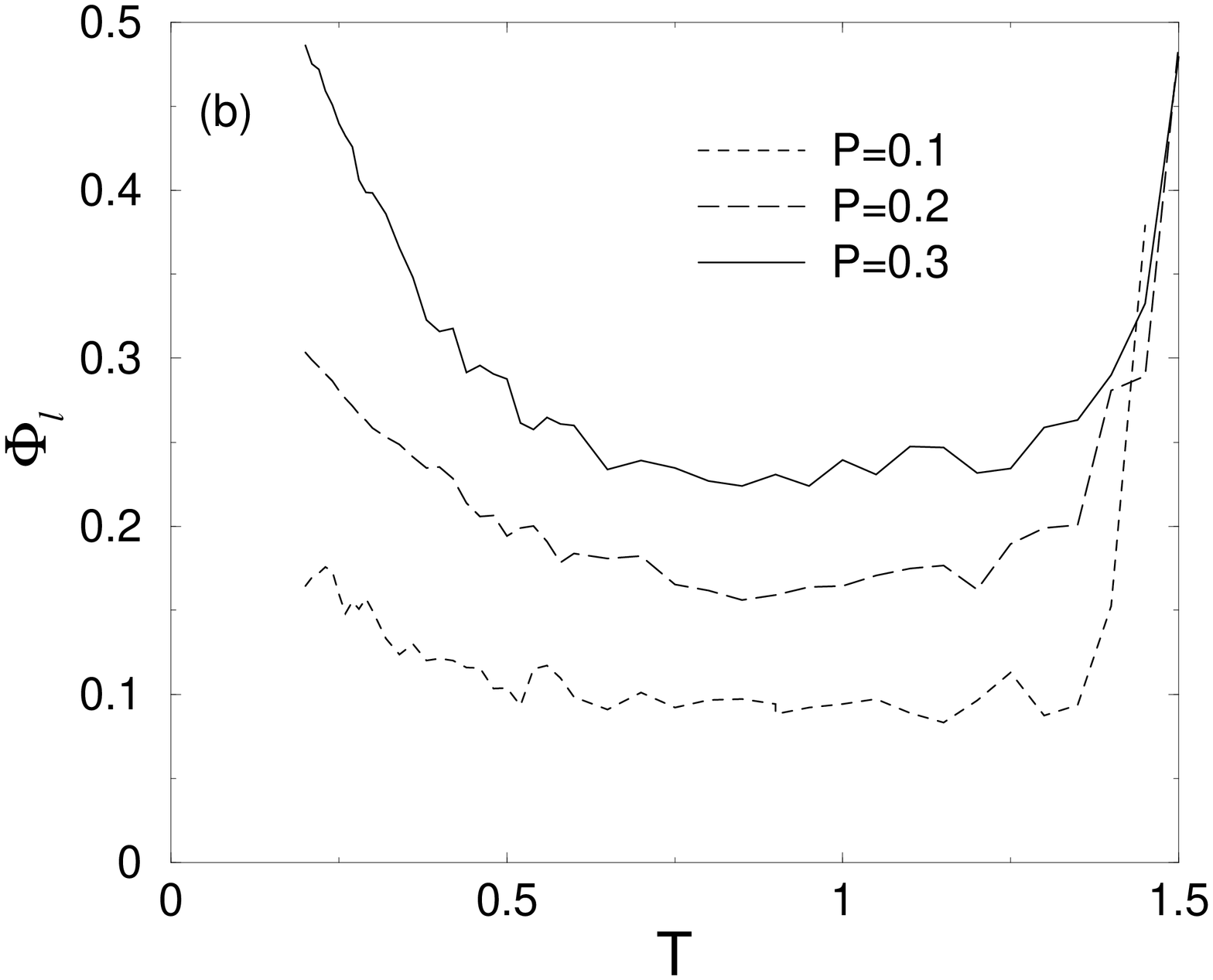}
}
\caption{(a) Symbols indicate our simulation results of the inverse
 Henry constant of hard sphere solutes in the Jagla liquid solvent for
 three different pressures. Lines indicate our theoretical predictions
 based on calculations of excess thermodynamic state functions---see
 Eq.~(\ref{eq:6x}) and Fig.~\ref{fig:DeltaG}.  Note the failure of the
 theoretical predictions for both high and low temperatures in the
 $P=0.3$ case caused by the fact that Eq.~(\ref{eq:6x}) is valid only in
 the limit of small liquid mole fraction $\Phi_\ell$ of solute in
 liquid. (b) The solubility of the hard sphere solutes in the Jagla
 solvent as function of temperature. The solubility minimum roughly
 coincides with the minimum of the inverse Henry constant.}
\label{fig:henry}
\end{figure}

Figure \ref{fig:henry}(a) shows the inverse Henry constant as a function
of $T$ for $P=0.1$, $0.2$, and $0.3$. The behavior of the solubility
follows the behavior of the inverse Henry constant
[Fig.~\ref{fig:henry}(b)], since for $T\ll T_{c_1}$ the partial vapor
pressure of solvent is very low and, therefore, $P_v \approx P$. Hence
the solubility $\Phi_\ell\approx P/k_H(T)$ is inversely proportional to
the Henry constant and thus the solubility minimum as function of
temperature practically coincides with the maximum of $k_H(T)$.  One can
see that Henry's law $\partial k_H(T,P)/\partial P\vert_T=0$ is
approximately correct since $k_H(T,P)$ increases by less than 30\% when
the pressure increases by 200\% from 0.1 to 0.3. The most interesting
feature of the behavior of the Henry constant is that it has a maximum
as function of temperature at $T_{\rm MH}=0.85\pm 0.10$, indicating that
the solubility of hard sphere solutes in a Jagla solvent has a minimum
at this temperature and increases upon cooling below $T_{\rm MH}$. This
behavior is similar to the behavior of the Henry constant of alkanes
(C$_n$H$_{2n+2}$) in water, which increases upon cooling below T=$373$K
\cite{Errington98}. Note that the $T_{\rm MH}$ is much larger than the
temperature of maximum density, $T_{\rm MD}=277$K, for both water and
for the Jagla model (for which $T_{\rm MD} \approx 0.5$ at $P=0.1$)
\cite{Xu06b}.

\section{Relation of the Henry constant to thermodynamic quantities}

To relate the behavior of the Henry constant to thermodynamic state
functions, we must express the Henry constant in terms of $\Delta g$.  At
low solute mole fraction $\Phi_\ell=y_\ell/(x_\ell+y_\ell)$ in the
liquid phase, where $x_\ell$ and $y_\ell$ are the number of moles of
solvent and solute respectively, the Gibbs potential $G(x_\ell,y_\ell)$
of the solution can be approximated by the first two terms in the
Taylor expansion of $G(x_\ell,y_\ell)$ around $y_\ell=0$ \cite{Huang}
\begin{equation}
G_\ell (x_\ell,y_\ell)=x_\ell g_\ell (0)+y_\ell\mu_\ell-y_\ell
T\Delta s_{\rm mix}(\Phi_\ell),
\label{eq:gl}
\end{equation}
where $g_\ell (0)$ is the molar Gibbs potential of pure liquid and
$\mu_\ell $ is the chemical potential of solute in liquid. At low
pressures, the analogous approximation can be made for the vapor phase
\begin{equation}
G_v(x_v,y_v)=x_v g_v(0)+y_v\mu_v -y_v T \Delta s_{\rm mix}(\Phi_v).
\end{equation}
Minimization of the sum of the Gibbs potentials of the two phases with
respect to $y_\ell$ at constant $y\equiv y_\ell+y_v$ gives the
equilibrium condition:
\begin{equation}
{\Phi_v\over\Phi_\ell }=\exp\left({\mu_\ell -\mu_v \over RT}\right).
\label{eq:Henry}
\end{equation}
The chemical potential of solute in vapor is approximately equal to the
molar Gibbs potential of the pure solute. Thus $\mu_\ell -\mu_v=\Delta
g_0$, where $\Delta g_0=\Delta g+T\Delta s_{\rm mix}$ and $\Delta g$ is
given by Eq.~(\ref{eq:DeltaG}). If we assume that the vapor is an ideal
gas, then $\mu_v=\mu_0(T)+RT\ln P$, and thus Eq.~(\ref{eq:Henry}) can be
rewritten in the form of the Henry law: $P_v/\Phi_\ell=k_H(T,P)$, where
$P_v=P\Phi_v$ is the partial pressure of solute in vapor and
\begin{equation}
k_H(T,P)=P\exp\left({\Delta g_0\over RT}\right)
\label{eq:6x}
\end{equation}
is the Henry constant which in the limit $P\to 0$ does not depend on
$P$. Thus the Henry constant has a maximum at the same temperature as
$\Delta g_0(T)/T$. Since the entropy of mixing does not depend on
temperature, this maximum coincides with the maximum of $\Delta g(T)/T$,
which as we have seen above corresponds to the temperature at which
$\Delta h=0$.

\begin{figure}[htb]
\centerline{ \includegraphics[width=12cm]{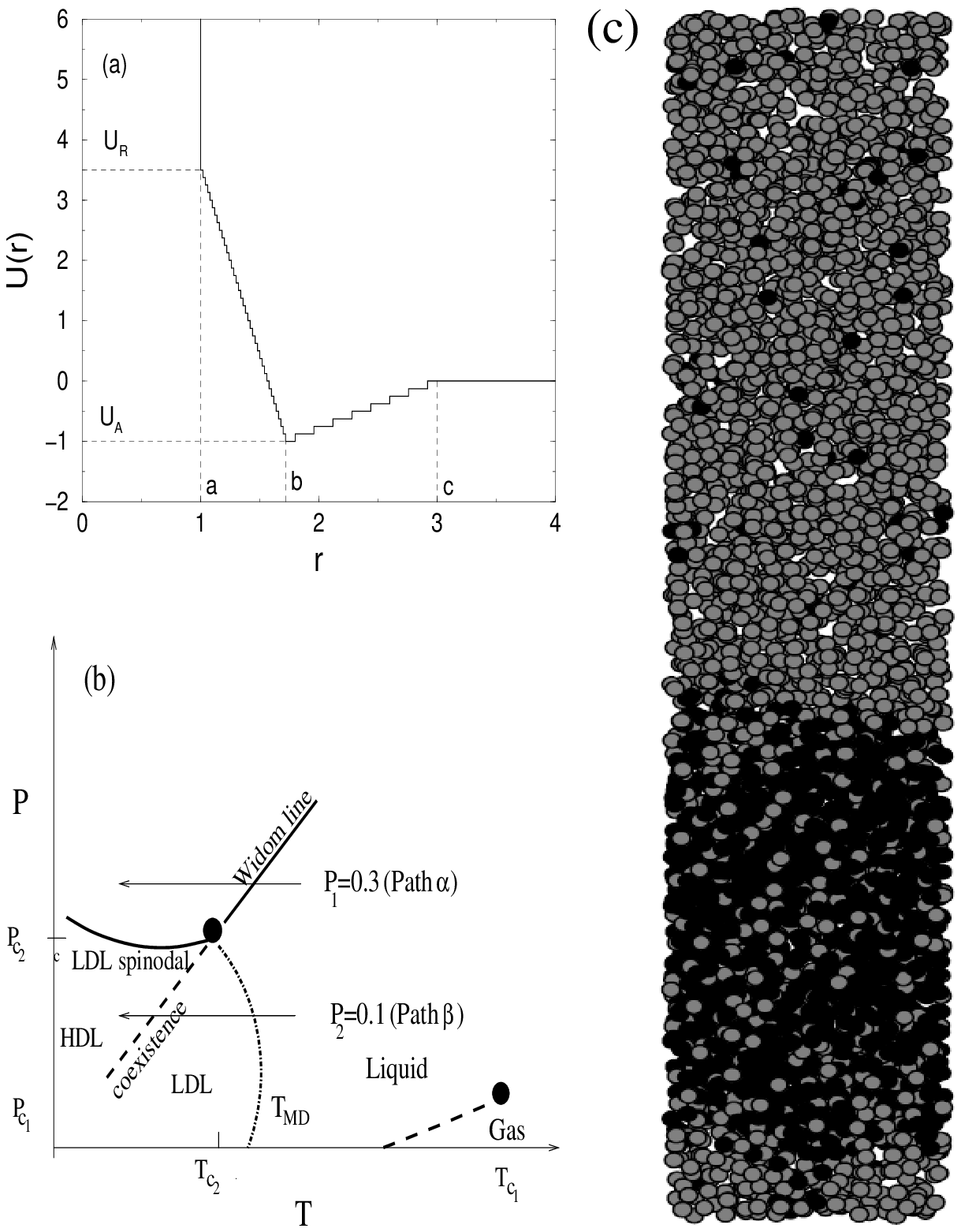}}
\singlespacing
\caption{(a) The discretized two-ramp Jagla potential captures much of
the essential physics corresponding to the first and the second shells
in water. The diameter of the hard core $r_1=a$ and the diameter of
the soft core $r_2=b\approx 1.72$ core determine the two essential
length scales corresponding to the first and second shells of
water. (b) Schematic P-T phase diagram of the two-ramp Jagla model of
plot ~\ref{fig:Jagla}(a). Shown are the two isobaric paths simulated,
Path $\alpha ~(P=P_1=0.3>P_{C_2})$ and Path $\beta
~(P=P_2=0.1<P_{C_2})$. Also shown are the liquid-gas and liquid-liquid
critical points (shown as solid circles), the corresponding
coexistence lines (dashed lines) and the Widom line (solid line). Also
shown is the locus of temperature of maximum density labeled $T_{MD}$
(dotted line). (c) A snapshot of a simulation box for calculation of
the Henry constant at $P_1=0.3>P_{C_2}$ and $T=0.95>T_{C_2}$. Black
circles represent the hard cores of the Jagla solvent, gray circles
represent hard sphere solutes.}
\label{fig:Jagla}
\end{figure}

To study the behavior of each thermodynamic term in
Eq.~(\ref{eq:DeltaG}) at high and low pressures, we compare a mixture of
Jagla particles and hard spheres with a given mole fraction $\Phi$ and a
given pressure $P$, with a pure solvent and pure solute at the same
pressures. We simulate the system at constant $P$ and $\Phi$ in a cubic
box with periodic boundaries with given numbers $N_S$ and $N_J$ of the
solute and Jagla particles, respectively, such that $N_J+N_S=1728$ and
$N_S/(N_J+N_S)=\Phi$. The systems were simulated at two pressures,
$P=P_1=0.3>P_{C_2}$ [path $\alpha$ of Fig.~\ref{fig:Jagla}(b)] and
$P=P_2=0.1<P_{C_2}$ [path $\beta$ of Fig.~\ref{fig:Jagla}(b)]. To find
the entropy of the system as function of $T$ and $P$, we change
temperature from $T=10$ to $T=0.3$ with a small step and perform
thermodynamic integration up to $T=10$ using the first two terms of the
virial expansion for $T>10$ with our analytically computed values of the
second virial coefficients for hard spheres and Jagla particles. We make
sure that the system does not phase segregate at any intermediate
temperature.

\begin{figure}[htb]
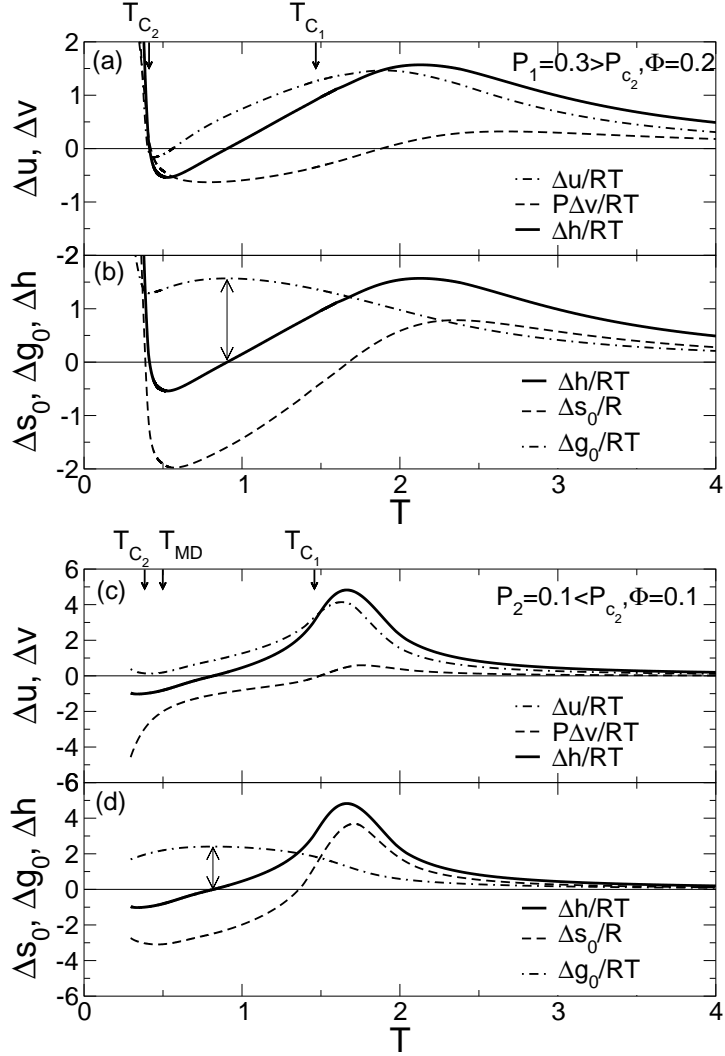

\begin{center}
\includegraphics[width=9.5cm]{fig4a.eps}
\includegraphics[width=9.5cm]{fig4b.eps}
\end{center}
\singlespacing
\caption{Excess thermodynamic state functions of hard sphere solutes in
 Jagla liquid solvent computed for (a,b) $P_1=0.3>P_{c_2}$ and
 $\Phi=0.2$ and (c,d) for $P_2=0.1<P_{c_2}$ and $\Phi=0.1$. One can see
 that the temperature at which $\Delta h=0$ corresponds to the maximum
 of $\Delta g/RT$ (arrow). Since $P\Delta v$ is only weakly
 $T$-dependent at high $P$, the decrease of $\Delta h=\Delta u+P\Delta
 v$ upon cooling for $P=0.3$ is mainly due to the decrease of the
 potential energy $\Delta u$, while at low pressure, the decrease of
 $\Delta v$ plays a major role. Note that as the system enters the
 region of negative thermal expansion coefficient of the pure solvent
 ($T<T_{\rm MD}$), $\Delta v$ for low $P$ becomes exceedingly negative
 corresponding to the simple physical picture that as the solvent is
 cooled, the solvent particles ``descend from the repulsive ramp''
 toward the potential minimum at $r_2=b$ so that the average distance
 between them increases, allowing more space for small solute particles
 to enter the solvent. At high pressure, both $\Delta u$ and $\Delta v$
 become positive at low $T$. This behavior is due to the rapid
 restructuring of the system as it enters the region of high density
 liquid above the positively sloped Widom line, which emanates from the
 critical point located at $T_{c2}=0.375$ and $P_{c2}=0.243$
 \cite{Xu06a}. This behavior leads to the rapid increase of $\Delta g_0$
 corresponding to the predicted decrease in solubility for $P=0.3$
 (Fig.~\ref{fig:henry}). However the calculated solubility in
 Fig.~\ref{fig:henry} does not decrease because, at this pressure, the
 actual solute mole fraction is so large that it is outside the region
 of applicability of Eq.~(\ref{eq:6x}).}
\label{fig:DeltaG}
\end{figure}

Figure~\ref{fig:DeltaG} shows the behavior of the dimensionless excess
state function $\Delta u/RT$, $P\Delta v/RT$, $\Delta h/RT$, $\Delta
s_0/R$, and $\Delta g_0/RT$. One can see that $\Delta h/RT$ becomes
negative at the temperature $T_{\rm MH}(\Phi, P)$ of maximal $\Delta
g_0/RT$ and hence, by Eq.~(\ref{eq:6x}), maximal Henry constant. We find
$T_{\rm MH}(\Phi=0.2,P=0.3)\approx0.90$ and $T_{\rm
MH}(\Phi=0.1,P=0.1)\approx 0.82$ thus only weakly depending on $\Phi$
and $P$. For both $P$ values, $\Delta v$ becomes negative as the
temperature drops below the liquid-gas critical temperature
$T_{c_1}=1.446$. Since $\Delta u$ decreases with decreasing temperature
in this region, $\Delta h$ becomes negative at sufficiently low $T$.

This behavior is consistent with the fact that the solute particles
enter relatively large free spaces between the solvent particles which
are kept apart not by the hard core but rather by the repulsive ramps,
thus creating large negative $\Delta v$ [Figs.~\ref{fig:Jagla}(b) and
\ref{fig:Jagla}(c)]. However at high temperatures, the entropic
contribution to the free energy is important and the solvent particles
in the solute are spread further apart than in the pure sovent. Thus the
number of solute particles in the attractive range of the Jagla
potential is reduced, leading to $\Delta u>0$.  As the temperature
decreases, the potential energy of the solution approaches its ground
state, which is the same as that of the pure solvent, because the solute
particles are small enough to remain in the area unavailable for the
solvent particles at low temperatures due to their repulsive ramp
interactions [Fig.~\ref{fig:Jagla}(b)].

At sufficiently low $T$, the excess enthalpy $\Delta h$ of the
solution becomes negative, indicating the increase of solubility upon
further cooling. Close to $T_{\rm MD}$, the behavior at low and high
pressure becomes different.  For $P=P_2<P_{c2}$
[Fig.~\ref{fig:DeltaG}(b)], the pure solvent starts to expand upon
cooling, thus increasing the free space for the solute particles, and
further increasing the solubility. This is clearly indicated by the
sharp decrease in $\Delta v$. For $P=P_1>P_{c2}$
[Fig.~\ref{fig:DeltaG}(a)], there is no $T_{\rm MD}$ and the pure
solvent collapses upon crossing the Widom line \cite{Xu06a} into an
HDL-like liquid on the lower side of the Widom line. Thus $\Delta v$
starts to increase upon cooling, and eventually $\Delta v$ becomes
positive. On the other hand $\Delta u$ keeps decreasing on decreasing
$T$ and becomes negative because the solute particles prevent the
solvent particles from ``climbing the repulsive ramps'' upon
compression and therefore help them remain near the minimum of the
pair potential.

\begin{figure}[htb]
\centerline{
\includegraphics[angle=270,width=10cm]{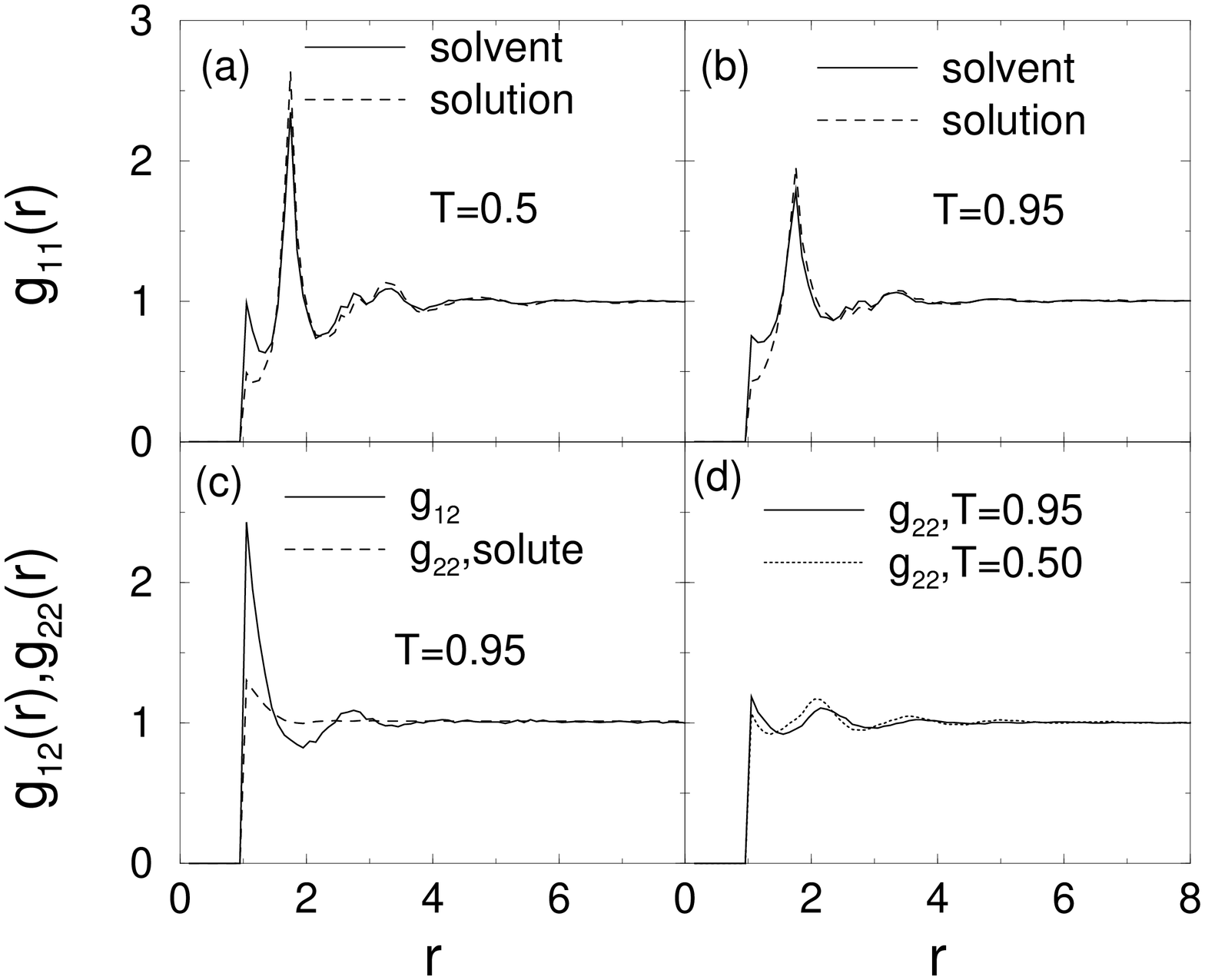}}
\centerline{
\includegraphics[angle=270,width=5cm]{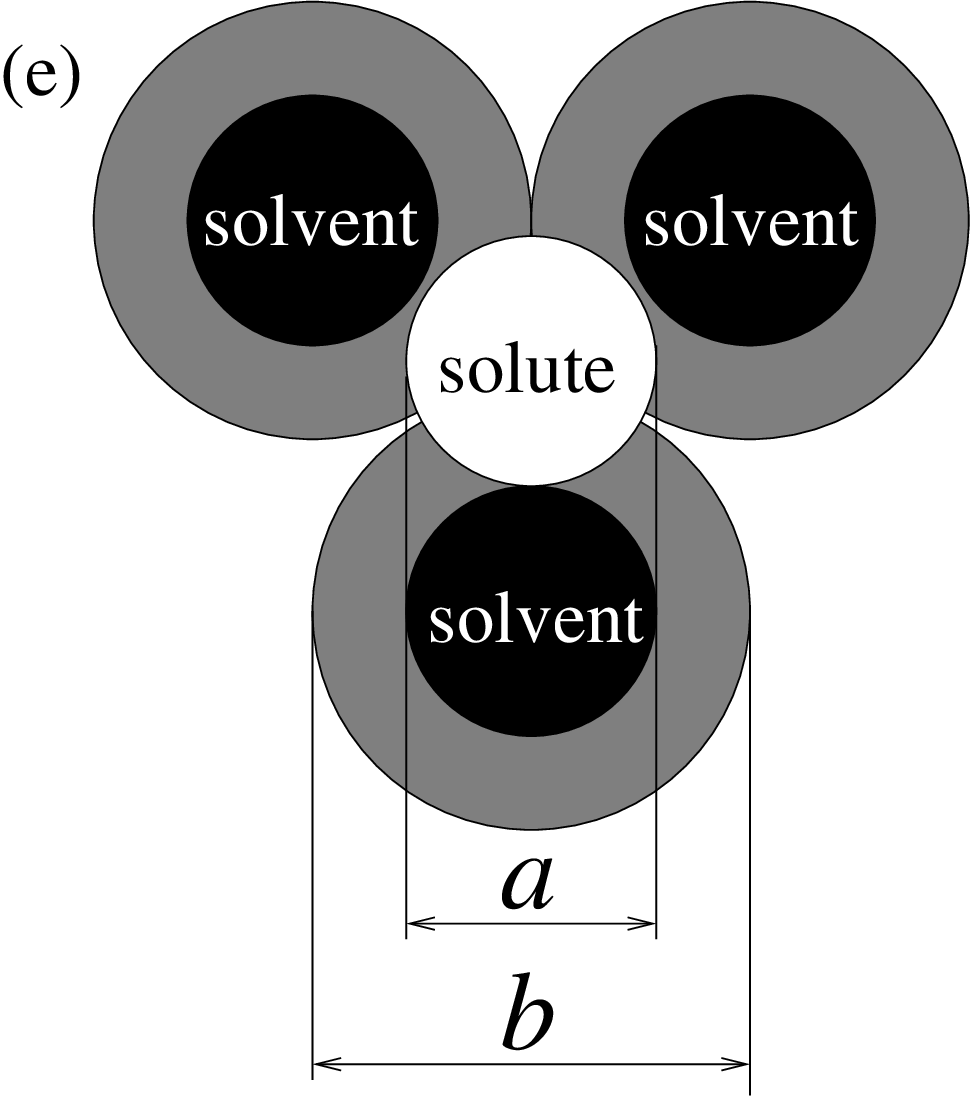}
\includegraphics[angle=270,width=4cm]{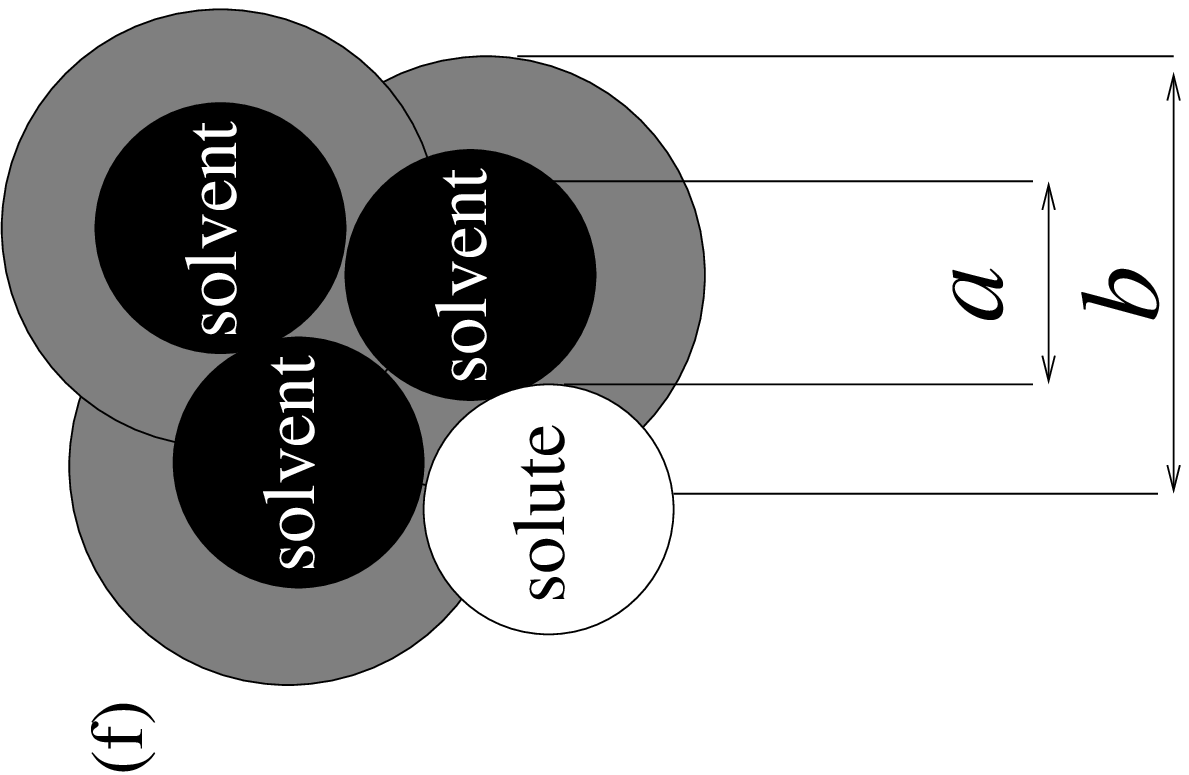}}

\caption{Solvent-solvent correlation function $g_{11}(r)$ computed at
 $P=0.3$ for pure Jagla solvent, and for the solution with mole
 fraction $\Phi=0.2$ at (a) $T=0.5$ and (b) $T=0.95$. These plots
 support the hypothesis that in solution the solvent particles prefer
 to stay at the soft core distance $r_2=b$ and not to ``climb the
 repulsive ramp'' occupied by the solute particles. (c) The
 solvent-solute correlation function $g_{12}$ indicates that the
 solute particles prefer to stay at the hard core distance $r_1=a$
 away from the solvent particles. For comparison we plot the
 correlation function of pure hard spheres at the same temperature
 $T=0.95$ and pressure $P=0.3$, showing a much smaller peak. (d) The
 solute-solute correlation functions $g_{22}(r)$ at two different
 temperatures and the same pressure $P=0.3$ show that at low
 temperatures the solute particles prefer to stay apart from each
 other, indicated by the decrease on cooling of the first peak and the
 increase of the second peak. This effect, that the solute particles
 clump less at low $T$, is in accord with the increase of solubility
 upon cooling (cf. Fig.~2). (e) Schematic of the Jagla solvent
 particles (black) surrounding a hard sphere nonpolar solute (white)
 of the same diameter $a$. At low $T$ the soft cores (gray) of the
 solvent particles do not overlap. The distance between the solvent
 particles corresponds to the potential energy minimum $r_1=b$ in
 Fig.~\ref{fig:Jagla}(a), thus the solvent particles form a cage
 around the hard sphere nonpolar solutes.(f) Schematic of the same
 situation but now for $T \approx T_{MH}$, the temperature of maximum
 Henry constant and hence minimum solubility.}
\label{fig:gr}
\end{figure}

To test this explanation directly, we compute for $P=0.3$ and $T=0.5$
the solvent-solvent density correlation function, $g_{11}(r)$, in the
pure solvent and in the solution with solute mole fraction $\Phi=0.2$,
[Fig.~\ref{fig:gr}(a)]. We see that the first peak in $g_{11}(r)$,
corresponding to the solvent particles staying at the hard core
distance in the pure solvent, significantly decreases in the
solution. In contrast, the second peak, corresponding to the solvent
particles staying at the soft core distance, increases. The behavior
of the cross-correlation function $g_{12}(r)$ of the solute and
solvent [Fig.~\ref{fig:gr}(c)] indicates that indeed the solute
particles prefer to remain very close to the solvent particles in the
region unavailable for solvent particles due to their repulsive ramp
interactions. The behavior of the solute-solute correlation function
$g_{22}(r)$ indicates a small decrease in the first peak and an
increase in the second peak upon cooling [Fig.~\ref{fig:gr}(d)],
corresponding to the increase of solubility. Overall, the structure of
the solution is more pronounced than in the pure solvent case,
indicated by negative $\Delta s_0$ [Fig.~\ref{fig:DeltaG}(b)].

This behavior of the Jagla solvent with hard sphere solutes is analogous
to the behavior of water molecules in the presence of nonpolar
compounds. The hard core of the Jagla model corresponds to the rigid
tetrahedron (first shell) of water molecules linked by hydrogen bonds to
a central molecule. A particle at the hard core distance in the Jagla
model corresponds to an extra (fifth) water molecule entering the first
shell. This molecule may cause the formation of a bifurcated hydrogen
bond \cite{Sciortino} of higher potential energy than the normal
hydrogen bond. The nonpolar solute molecules in water remain close to
the first shell of water molecules, and thus prevent extra water
molecules entering into the first shell and forming bifurcated hydrogen
bonds. Thus in water, solute particles help stabilize hydrogen
bonds. The second shell of water becomes more structured by forming
cages around the solute molecules without breaking any hydrogen bonds.

%but instead by increasing their average strength.

For $P=P_2<P_{c_2}$, the $g_{11}(r)$ for pure solvent and solution are
practically indistinguishable (not shown), with only slight decrease of
the overall number of solvent particles within the attractive range in
the solution compared to the pure solvent, corresponding to small
$\Delta u>0$. The behaviors of $g_{12}(r)$ and $g_{22}(r)$ at low
pressures (not shown) are practically the same as at high
pressures. This means that for low pressures the increase of the
solubility upon cooling is mainly because of the decrease of $\Delta v$
due to the small and even negative thermal expansion coefficient of the
pure solvent. We expect that the relative strength of the two
contributions, energetic and volumetric, to $\Delta h$ strongly depends
on the size of the solute particles. Accordingly, in water one might
expect both mechanisms to be present, depending on $P$, and on the nature
of the solute.

To check the validity of our calculations of the Henry constant $k_{\rm
H}$ and the thermodynamic excess quantities, we compare $k_{\rm H}$
obtained directly by measuring the mole fraction of hard spheres in the
slabs of Sec.~III with $k_{\rm H}$ computed using
Eq.~(\ref{eq:6x}). Figure \ref{fig:henry} shows that although the
individual contributions of different terms to $\Delta g$ behave in a
complex way, the resulting behavior of the Henry constant is in good
agreement with the direct simulations near the solubility minimum. Note
the failure of thermodynamic predictions at high and low
temperatures. The dramatic solubility drop predicted by
Eq.~(\ref{eq:6x}) at $P=0.3$ is because at fixed mole fraction, the
amount of solute particles become insufficient to prevent the solute
from collapse, and eventually $\Delta u$ also starts to increase upon
cooling making $\Delta h >0$ at low temperatures. This predicted
decrease of the solubility upon cooling is not observed in the direct
simulations (Fig.~\ref{fig:henry}). The discrepancy between the
thermodynamic predictions and actual behavior of the Henry constant is
due to the fact that at low temperatures the mole fraction $\Phi_\ell >
0.3$ is so high, that the linear Taylor expansion of Eq.~(\ref{eq:gl})
is no longer valid. Thus, in reality, there are still enough solute
particles to prevent the solution from collapsing into an HDL-like
liquid. In the next section, concerning polymer solution, we observe the
decrease of solvent quality upon cooling at $P>P_{c_2}$ because the
polymer monomers do not have sufficient local mole fraction to keep the
solution from collapse.

\section{Polymer swelling upon cooling}

In order to relate the maximum of solubility of nonpolar compounds to
the polymer LCST and protein denaturation, we next study the behavior of
a polymer composed of $M=44$ monomers modeled by hard spheres of
diameter $d_0=a$. We model covalent bonds by linking the hard spheres
with the potential
\begin{equation}
U_{\rm bond}(r) = \left\{
\begin{array}{ll}
\infty & r < d_1\\ 0 & d_1 < r <d_2 \\ \infty & r > d_2
\end{array}\right. ,
\label{eq:bond}
\end{equation}
where the minimal extent of the bond is $d_1=a$ and the maximal
extent is $d_2=1.2a$.  We simulate for $t_{m}$ time units the trajectory
of the polymer at constant $T$ and $P$ in a cubic box containing
$N=1728$ Jagla solvent particles with periodic boundaries.  We calculate
the polymer radius of gyration $R_g(t_k)$, where
\begin{equation}
R_g^2(t_k)\equiv {M\sum_{i=1}^M (x_i^2 +y_i^2 +z_i^2)
  -\left(\sum_{i=1}^M x_i\right)^2 -\left(\sum_{i=1}^M x_i\right)^2
  -\left(\sum_{i=1}^M x_i\right)^2\over M^2}
\end{equation}
at equidistant times $t_k\equiv k\Delta t$, where $\Delta t\equiv 1$, and we
calculate the rms radius of gyration $R_g$, where $R_g^2\equiv\sum_{k=1}^m
R_g^2(t_k)/m$, and $m\equiv t_m/\Delta t$.

\begin{figure}[htb]
\centerline{\includegraphics[angle=270,width=16cm]{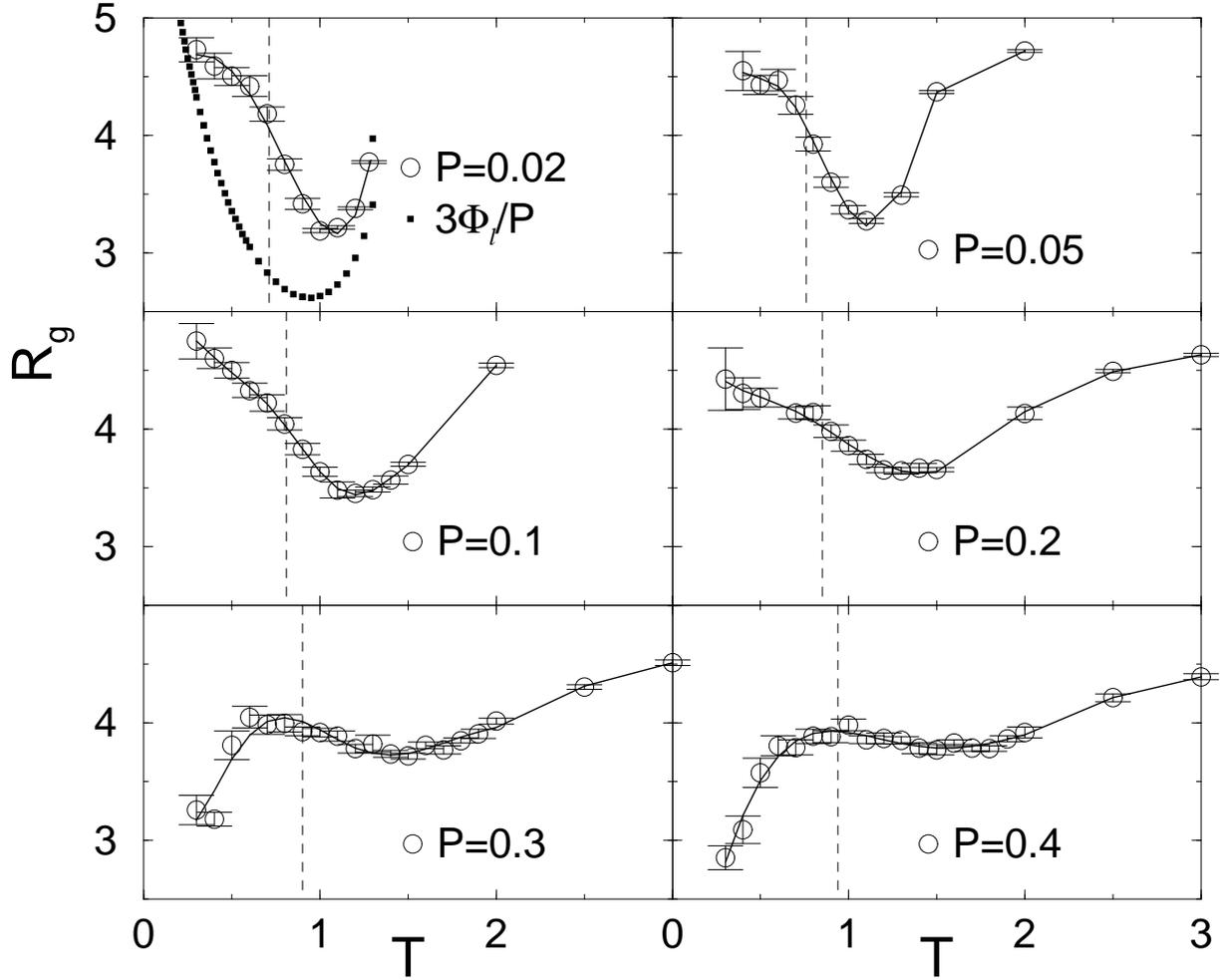}}
\caption{Dependence on temperature of the average radius of gyration of
$M=44$ polymer comprising $M=44$ hard sphere monomers for different
pressures. The dashed vertical lines show $T_{\rm MH}(P)$ of the maximum
of the Henry constant. We also show that for the lowest pressures the
solubility minimum of the hard sphere solute particles roughly coincides
with the $R_g$ minimum.  We find the rms radius of gyration in a vacuum
is $R_{gv}=4.77$. The evolution of the polymer chain in the solvent is
much slower than in a vacuum, so to calculate a reliable $R_g$, we must
carefully estimate the error bars. In order to do so we divide the
entire set of $m$ measurements into $m/n$ equal groups of $n\leq m/2$
subsequent values $R_g^2(t_k)$, and find the average for each group
$\langle R_g^2\rangle_i$, $i=1,2,...,m/n$ and the standard deviation
$\sigma_n$ of $\langle R_g^2\rangle_i$. If the values $\langle
R_g^2\rangle_i$ were independent, the error bars on their average could
be determined as $\sigma_n/\sqrt{m/n-1}$, and this estimate would not
depend on $n$. In fact, for small $n$, $\sigma_n/\sqrt{m/n-1}$ is an
increasing function of $n$, which reaches the plateau at $n= \tau/\Delta
t$, where $\tau$ is the correlation time. Accordingly, we determine the
error bar as $\max_n[\sigma_n/\sqrt{m/n-1}]$, checking that this
quantity indeed reaches a plateau for a given number of observations
$m$. If this quantity does not reach the plateau, it means that the time
of averaging is insufficient and we must increase $m$.  }
\label{RG}
\end{figure}

\begin{figure}[htb]
\centerline{\includegraphics[width=12cm]{Rg-map.eps}
}
\centerline{\includegraphics[width=10.5cm]{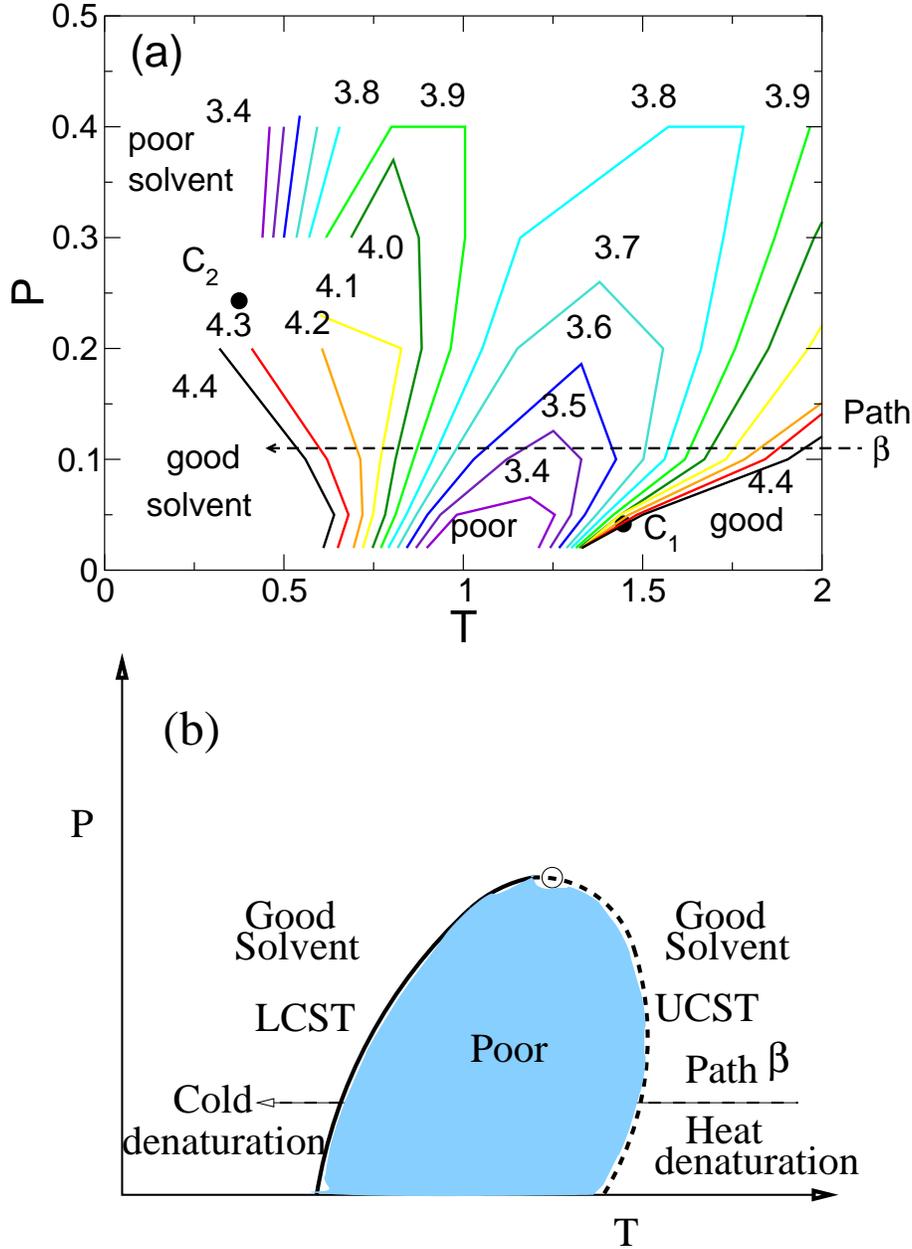}}
\singlespacing
\caption{ (a) Contours of equal $R_g$ in the P-T plane, showing
regions of good and poor solvent; the numbers denote the value of
$R_g$. The filled circles $C_1$ and $C_2$ indicate the liquid-gas and
liquid-liquid critical points. Note that at low $P$, on decreasing $T$
along path $\beta$, one passes from a region of good solvent (swollen
``denatured'' polymers) to a region of poor solvent (collapsed
polymers) and finally to a region of good solvent (``cold
denaturation''). (b) Schematic illustration of the loci of the lower
critical solution temperature (LCST) and the upper critical solution
temperature (UCST) for a polymer chain such as studied here. The UCST
delineates the region of high-$T$ swelling (``heat denaturation of a
protein'') while LCST delineates the region of low-$T$ swelling
(``cold denaturation of a protein'') sampled by path $\beta$.}
\label{RG35}
\end{figure}

Figure \ref{RG} shows the behavior of the radius of gyration of a
polymer as a function of temperature for six values of pressure. Also
shown for the $P=0.02$ case is the inverse Henry constant. One can see
that for small pressures, $R_g$ reaches a deep minimum at $T=T_{\rm
MR}(P)$, which occurs above the maximum of the Henry constant $T=T_{\rm
MH}(P)$. As pressure increases, the minimum becomes less pronounced and
$T_{\rm MR}(P)$ shifts to higher temperatures and eventually, at
$P=0.4$, almost disappears. According to the de~Gennes-Flory-Huggins
theory \cite{Rubinstein03}, the radius of gyration is a function of
polymer length and the solvent quality $\chi$:
\begin{equation}
R_g=\ell N^{1/2} f(z),
\end{equation}
where $z=(2\chi-1)N^{1/2}$ and
\begin{equation}
f(z) \sim \left\{
\begin{array}{ll}
|z|^{-1/3} & z \ll 0 \\ 1 & z=0,\\ z^{2\nu-1} & z>0
\end{array}\right.,
\label{eq:Rgz}
\end{equation}
where $\nu \approx 0.588$ is the Flory exponent for $R_g$. Values of
$\chi <1/2$ correspond to a ``poor'' solvent, in which the polymer chain
is collapsed and $R_g \sim N^{1/3}$. Values of $\chi>1/2$ correspond to
a ``good'' solvent where the polymer swells and $R_g \sim N^{\nu}$. The
value $\chi =1/2$ corresponds to the ``theta condition'' with $R_g \sim
N^{1/2}$. Thus for a polymer of a given length, the radius of gyration
monotonically increases with solvent quality and can be used to
calculate lines of the equal solvent quality in the $P-T$ plane. We
tested numerically that near the minimum of $R_g$ for $P\leq 0.2$, $R_g
\sim N^{1/3}$, so the solvent quality is poor. For $T\leq 0.5$, $R_g$ is
comparable with $R_{gv}=4.77$ in a vacuum, and hence the solvent quality
is good.  Thus the theta condition corresponds to some temperature
between $T=0.5$ and the value of $T$ at which $R_g$ has a
minimum. Accordingly, the LCST which differs from the theta point by a
term of the order of $1/\sqrt{M}$ \cite{Rubinstein03} is also located in
this region. Assuming that $R_g=3.5$ for $N=44$ corresponds to the theta
point, we construct a region in the $P-T$ plane of approximate location
of the LCST which merges with the upper critical solution temperature
(UCST) at $P=0.12$ and $T=1.25$ (Fig.~\ref{RG35}).  The region of the
$P-T$ plane below this line corresponds to the poor solvent in which
sufficiently long polymer chains segregate from the sufficiently
concentrated solution. This region lies below $P_{c_2}$ and
significantly above the $T_{\rm MD}$ line in the $P-T$ plane. On the
high temperature side, the region where $R_g<3.5$ is bounded by the
liquid-gas critical point.

Since the driving force of the folding of globular proteins is the
formation of the hydrophobic core, one can imagine that the regions in
the $P-T$ plane in which the various proteins can fold into their native
states must have similar shapes.

\section{Discussion and Conclusions}

Our work suggests a physical mechanism for the increase of solubility of
nonpolar solvents upon cooling (Fig.~\ref{fig:henry}) connecting the
cold denaturation of a polymer to the LCST. These phenomena are related
to thermodynamic anomalies of pure water because both thermodynamic and
solubility properties are reproduced in the simple Jagla solvent
model. Both of these are caused by the existence of two repulsive scales
in the effective model potential, the hard core corresponding to the
first tetrahedral shell of water molecules and the soft core
corresponding to the second shell of water molecules. However, the
temperature range in which cold denaturation and the solubility minimum
occur does not directly coincide with the regions of other water
anomalies, which are generally placed at lower temperatures.

Interestingly, the region of poor solvent, corresponding to the sharp
minimum in $R_g$ lies below the critical pressure of the second critical
point. Above this pressure, the behavior of the solvent quality changes,
being almost constant in the wide range of temperatures much above
$T_{c2}$, but dramatically decreasing as we approach $T_{c2}$,
indicating that the HDL in the Jagla model is a poor solvent, while the
LDL is a good solvent.  This behavior is in accord with the
thermodynamic studies of Section IV where a sharp increase in $\Delta g$
upon cooling near the Widom line is found.  The physical reason for this
is that LDL in the Jagla model has a larger free volume for the hard
spheres to enter and is also less ordered than HDL, so entering of the
hard spheres does not destroy this order. Whether this situation for
water is the same as or the opposite is not clear, since LDL in water,
although having larger volume, is more ordered than HDL. New
simulations, underway, involving water models that have a liquid-liquid
critical point in the accessible region and a chain of Lennard-Jones
monomers may answer this question.

\subsection*{Acknowledgments}

We thank C. A. Angell, P. G. Debenedetti, M. Marqu\'es, P. J. Rossky,
S. Sastry, and Z. Yan for helpful discussions, the Office of Academic
Affairs of Yeshiva University for sponsoring the high performance
cluster, and NSF for financial support.

\newpage

\newpage

\newpage

\newpage

\newpage

\end{document}